# Group delay dispersion tuned femtosecond Kerr-lens mode-locked Ti:sapphire laser

E. Seres*, J. Seres, T. Schumm

Atominstitut - E141, Technische Universität Wien, Stadionallee 2, 1020 Vienna, Austria

*enikoe.judit.seres@tuwien.ac.at

**Abstract**

We report on a new design for a femtosecond Ti:sapphire oscillator in which dispersion compensation is realized exclusively using mirrors, including special mirrors with third order dispersion. This makes the oscillator dynamically tunable over a spectral range of 45 nm using an intracavity wedge-pair; and it delivers 40 fs pulses at 80 MHz repetition rate. Due to the all-mirror design, the oscillator represents an attractive base for a tunable frequency comb for high precision spectroscopy applications.

**Introduction**

Utilizing the broad gain bandwidth of Ti:sapphire crystals, pulsed oscillators tunable over broad spectral range [1-6] were realized or ultrashort pulse durations down to 10 fs [7, 8] and far below [9-12] were reached. These properties made these lasers attractive sources for spectroscopy, sensing, non-linear microscopy, coherent Raman imaging and coherence tomography applications. Furthermore, studying ultrafast processes with high temporal resolution and building frequency combs for high precision spectroscopy became possible.

Pulsed oscillators require suitable, low loss optical elements with negative intracavity group delay dispersion (GDD) to compensate the positive GDD of the used laser crystal and additional intracavity optical elements. Broadband negative GDD multilayer mirrors are one of the possible solutions. Such oscillators support the mentioned ultrashort pulse durations; they are not sensitive to short- and long-term beam directional stability and consequently they are excellent candidates to realize frequency combs [13] with repetition rate and carrier envelope phase (CEP) stabilization. They only drawback so-far was that they were able to work only on a particular central wavelength, and they were not tunable. This study shows how to overcome this limitation.

Usually, to realize tunable laser oscillators, Brewster prism-pair based compressors are placed into the cavity, which produces the necessary negative GDD and at the same time, serves as a spectral filter to select and tune the central wavelength by an aperture slit. With such oscillators, tuning ranges up to 100-400 nm were reached [1, 3, 5, 6] at 50-150 MHz repetition rates with pulse durations below 100 fs. At a low repetition rate of 22 MHz, 170 nm tuning range and 150-300 fs pulse duration was demonstrated [4].

Other realizations utilized birefringent filters together with specially designed broadband chirped mirrors [2] and an oscillator with 70-fs-long pulses was realized with a tunability in the 800 nm to 900 nm spectral range. Synchronously pumped [14] or actively mode-locked [15] oscillators can be tuned by exploiting that the first order dispersion, or group delay and consequently the round-trip time, is wavelength dependent. Changing the cavity length, the wavelength was forced to adapt to get the round-trip time synchronized with the external source. This method cannot be used for continuous wave (CW) pumped passively mode-locked oscillators, which simply change their repetition rate by the change of the cavity length and their wavelength remains unaltered.

We have developed a new and simple, continuously tunable, mirror based Kerr-lens mode-locked Ti:sapphire oscillator. The design is based on a mirror pair having negative GDD and at the same time positive TOD (third order dispersion). The oscillator is continuously tunable via the positive GDD of a wedge-pair and operates at the wavelength where the intracavity GDD is near zero. In the present configuration, it is tunable in the 750-795 nm range, which is the relevant range for our planned spectroscopic applications, the excitation of the [229]Thorium low-energy isomeric nuclear state, using the 5[th] harmonic in the vacuum ultraviolet regime [16, 17].

**Oscillator setup**

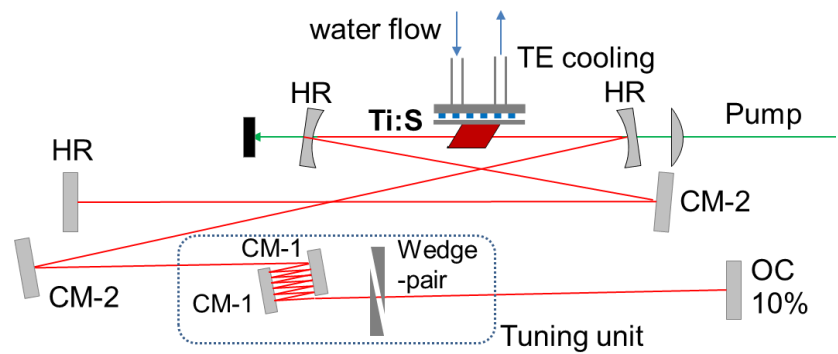

Fig. 1: Oscillator arrangement. CM: chirped-mirror; OC: output coupler; HR: high-reflector mirror; Ti:S: Ti:sapphire crystal.

The layout of the oscillator is illustrated in Fig. 1. It is a mirror-based oscillator setup with a cavity design suggested in [8, 18]. The Ti:sapphire crystal is 3.6 mm long with 80% absorption at the 532 nm wavelength of the pump laser (finesse pure 16, Laser Quantum). The two curved mirrors around the crystal have focal lengths of 50 mm. The two CM-2 are usual broadband chirped-mirrors (Layertec) with -40 $fs^2$ GDD and low TOD. The tunability of the oscillator is given by the CM-1 special mirror-pair (10Q20UF.40, Newport) with large negative GDD and large positive TOD. In the spectral range between 750 nm and 800 nm, its GDD changes from -40 $fs^2$ to -80 $fs^2$ meaning a TOD of about +760 $fs^3$. Together with the BK7 wedge-pair, it provides a continuously tunable GDD with slightly changing TOD inside the cavity. The oscillator cavity is designed with short/long arm length of about 770/1100 mm to set the repetition rate to 80 MHz. The design offers the possibility to turn the oscillator into a frequency comb by future repetition rate and CEP stabilization. Furthermore, the tuning range of 750-795 nm is suitable to generate VUV pulses in the spectral range of 150-159 nm for future spectroscopic application.

**Working principle of tunability**

Earlier dispersion tuning methods, as mentioned above, rely on the fact that the cavity round-trip time or the group delay GD($\omega$) of the cavity is spectrally dependent, and the laser becomes tunable by changing the cavity length. Or scheme is based on the spectral dependence of the intracavity group delay dispersion GDD($\omega$) due to a third-order dispersion TOD:

$$GDD(\omega, d) = GDD_0 + GVD_0 \cdot d + \frac{1}{3}TOD_0(\omega - \omega_0) \tag{1}$$

where the index 0 refers to the quantities at $\lambda_0$ = 800 nm or $\omega_0$ = 750π THz. $GDD_0$ is the cavity GDD without the wedge-pair, $GVD_0$ = 44.6 fs$^2$/mm is the group velocity dispersion of the BK7 material of the wedge-pair and $d$ is the thickness of the glass inserted by the wedge-pair. $TOD_0$ is the intracavity TOD mainly caused by the multiple reflections on the CM-1 mirrors. Using 12 reflections, it is in the range of 9000 fs$^3$ and the small 151 fs$^3$ and 32 fs$^3$/mm contributions of the sapphire crystal and the BK7 wedge-pair respectively can be neglected. A Kerr-lens mode-locked oscillator can run at two different modes: (i) At the wavelength of maximal gain of the cavity. Here, the pulse duration (spectral width) changes by changing the intracavity GDD [8, 19]. (ii) When the effect of the Kerr-lens is maximal. According to our knowledge, this mode has not been demonstrated yet. At this mode, the pulse is shortest, and the wavelength is determined by realizing an intracavity

$$GDD(\omega, d) \approx 0 . \tag{2}$$

In fact, the intracavity GDD should be slightly negative to compensate the effect of the self-phase modulation (SPM) of the mode-locking in the laser crystal [8]. Eq. (2) can be fulfilled with suitable $\omega$-$d$ pairs. This second mode was realized in our scheme by applying larger pump power (3.8 W instead of 3.5 W) and altering the cavity (the distance between the two curved mirrors) further away from the CW optimal position. At a moderate tuning range, < 10% of the central wavelength, the approximation $\lambda_0^{-1} - \lambda^{-1} \approx (\lambda - \lambda_0)/\lambda_0^2$ can be applied to get the operational wavelength:

$$\lambda \approx \lambda_0' + \frac{3\lambda_0^2 \cdot GVD_0}{2\pi c \cdot TOD_0} d, \quad \text{with} \quad \lambda_0' = \lambda_0 + \frac{3\lambda_0^2 \cdot GDD_0}{2\pi c \cdot TOD_0}, \tag{3}$$

which means an approximately linear dependence of the operational wavelength to the thickness $d$ of the inserted glass of the wedge-pair.

**Experimental results**

The described oscillator was realized using 12 reflections on the CM-1 mirror-pair and inserting different amounts of BK7 glass into the cavity. The oscillator was continuously tunable within a broad spectral range, between 760 nm and 795 nm as plotted in Fig. 2(a) together with measured spectra at 4 central wavelengths. Operation at 750 nm was realized by removing the wedge-pair and the generated spectrum is also plotted in Fig. 2(a) and the corresponding measurement points in Fig. 2(b). Additional peaks are observable at the pedestals of the spectra, which appear when the phase matching between dispersive wave, solitary wave and periodic perturbation is fulfilled [8]. As can be seen in Fig. 2(a) and plotted separately in Fig. 2(b), the central wavelength of the spectra is continuously shifted near-linearly with the thickness of the inserted BK7 glass, as expected from Eq. (3), with a slope of about 4.7 nm/mm. The spectral width changed from 21 nm to 25 nm, plotted in Fig. 2(b), corresponding to a transform limited pulse duration between 43 fs and 36 fs, respectively. To check the pulse durations, autocorrelation curves were measured directly at the oscillator output at 760 nm and 795 nm central wavelengths and presented in Fig. 2(c) and Fig. 2(d). At 760 nm, the autocorrelation curve shows a transform limited 44-fs-long pulse, while at 795 nm, the pulse was near to the transform limit. The measured pulse duration was 40 fs, which was about 10% longer than the transform limit. The corresponding small chirp can be easily compensated extra-cavity. The oscillator was pumped with 3.8 W power in all measurements and the output power changed slightly from 0.50 W to 0.52 W while tuning from 760 nm to 795 nm.

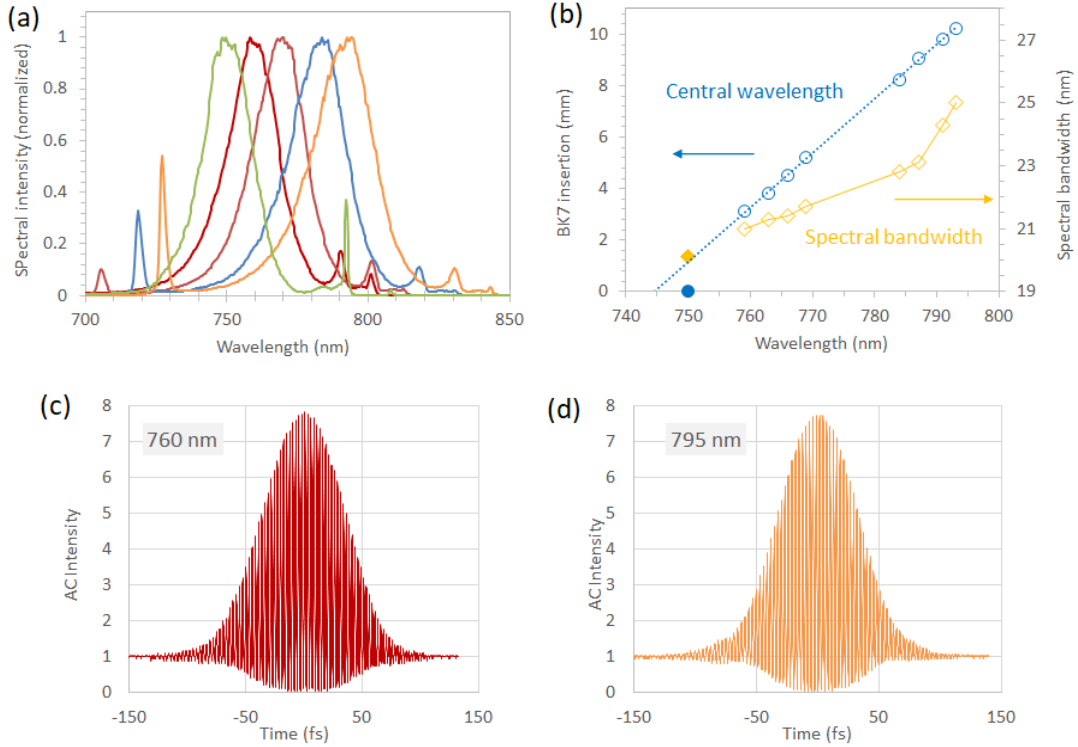

Fig. 2: (a) Measured spectra tuned over the 750 nm to 795 nm spectral range. (b) The central wavelength of the spectra has linear dependence on the inserted thickness of the intracavity BK7 glass (wedge-pair). The spectral width also changed by the wavelength tuning between 21 nm and 25 nm. (c) and (d) are the measured autocorrelation curves at central wavelength of 760 nm and 795 nm, respectively.

## Discussion

As mentioned above and depicted in Fig. 2(b), the operational wavelength depends linearly on the inserted BK7 glass thickness, as described by Eq. (3). The fitted trend line yields $\lambda'_0$ = 745 nm and a slope of 4.7 nm/mm. According to these values, the laser should operate at 745 nm when the wedge-pair is removed. Actually, it operates at 750 nm (or slightly below, at 748 nm) as indicated in Fig. 2(b). The reflectance of the CM-1 mirrors starts to decrease at 750 nm, which impedes the oscillator to operate much below that wavelength.

Using the values of the fitted trendline in Eq. (3), it is possible to calculate the $GDD_0$ and the $TOD_0$ of the CM-1 mirror-pair at $\lambda_0$ = 800 nm. After correcting with the TOD of the sapphire and the ≈11 mm BK7 glass, the settings necessary for the operation at 800 nm, the calculation yields 758 fs³/bounce for the TOD. $GDD_0$ means without BK7 glass, however the GDD of the laser crystal, air, and additional chirped mirrors are considered, gaining -57 fs²/bounce. While the TOD yields the expected value, the GDD was found to be somewhat smaller. However, if one expects that the dispersion of the mirrors follows the specifications, the cavity has to be at -270 fs² dispersion to form a stable soliton. In that case, the side-peaks in the spectra should appear +45 nm and -65 nm from the central wavelength. In the measurements, the positions of the side-peaks were +35 nm and -65 nm, very close to the expected values. The small difference is attributed to the manufacturing inaccuracy of the mirrors and the effect of higher order dispersions.

**Summary**

A continuously tunable Ti:sapphie oscillator has been developed using the intracavity GDD and the large TOD of a special mirror-pair as basic tuning principle. The oscillator cavity was built using only mirrors and a thin wedge-pair. Short pulses of 40 ± 4 fs pulses were obtained in the spectral range of operation between 750 nm and 795 nm, at a repetition rate of 80 MHz. The oscillator can be tuned to longer wavelength by adding further dispersive material to the cavity; however, it is practical to use less reflections on the mirror pair instead, if operation at longer wavelength range is required.

The demonstrated oscillator is suited for converting its wavelength by high harmonic generation using gas or solid medium, which has been demonstrated at 800 nm with extra-cavity [20, 21], enhancement cavity [22, 23] and intra-oscillator [24] arrangements. In that way, a tunable, short pulse laser source can be obtained in the ultraviolet spectral range by $3^{rd}$ harmonic generation or in the vacuum ultraviolet spectral range between 150 nm and 159 nm by $5^{th}$ harmonic generation for different spectroscopic applications. Such a spectral range in the vacuum ultraviolet is of special interest, as it corresponds to the predicted excitation wavelength of the $^{229}$Thorium isomeric nuclear state [16, 17].

The required moderate 3.8 W pump power opens the way for further development in the direction of increased output power and a suitable dispersion design for obtaining shorter pulse durations and an even broader tuning range.

**Funding**

This project has received funding from the European Research Council (ERC) under the European Union's Horizon 2020 research and innovation program (grant agreement n° 856415). This project 20FUN01 TSCAC has received funding from the EMPIR program co-financed by the Participating States and from the European Union's Horizon 2020 research and innovation program.